\documentclass[unsortedaddress, pre, twocolumn, showpacs, amsmath, amssymb, superscriptaddress]{revtex4-1}

\usepackage{subfigure}
\usepackage{bm}
\usepackage{multirow}
\usepackage{bbm}
\usepackage[utf8]{inputenc}
\usepackage[T1]{fontenc}
\usepackage{graphicx}
\usepackage[english]{babel}
\usepackage{pdflscape}
\usepackage{natbib}
\usepackage{times}
\usepackage{ae}
\usepackage{dcolumn}
\usepackage{tikz}

\newcommand{\f}{\frac}

\newcommand{\be}{\begin{equation}}
\newcommand{\ee}{\end{equation}}
\newcommand{\ba}{\begin{array}}
\newcommand{\ea}{\end{array}}
\newcommand{\bc}{\begin{center}}
\newcommand{\ec}{\end{center}}

\begin{document}

\title{Inertia effects and stress accumulation in a constricted duct: \\ A combined experimental and lattice Boltzmann study}

\date{\today}

\author{T.~Kr\"uger}
\email{t.krueger@mpie.de}
\affiliation{Max-Planck-Institut f\"ur Eisenforschung, Max-Planck-Str.~1, 40237~D\"usseldorf, Germany}
\author{M.~A.~Fallah}
\affiliation{Department of Experimental Physics I, University of Augsburg, Universit\"atsstr.~1, 86159~Augsburg, Germany}
\author{F.~Varnik}
\affiliation{Interdisciplinary Center for Advanced Materials Simulation, Stiepeler Str.\ 129, 44780 Bochum, Germany}
\affiliation{Max-Planck-Institut f\"ur Eisenforschung, Max-Planck-Str.~1, 40237~D\"usseldorf, Germany}
\author{M.~F.~Schneider}
\affiliation{Department of Mechanical Engineering, Boston University, 110~Cummington~St., Boston, MA, 02215, USA}
\author{D.~Raabe}
\affiliation{Max-Planck-Institut f\"ur Eisenforschung, Max-Planck-Str.~1, 40237~D\"usseldorf, Germany}
\author{A.~Wixforth}
\affiliation{Department of Experimental Physics I, University of Augsburg, Universit\"atsstr.~1, 86159~Augsburg, Germany}

\pacs{47.11.-j, 47.15.-x, 47.80.-v, 47.63.Cb}

\begin{abstract}
We experimentally and numerically investigate the flow of a Newtonian fluid through a constricted geometry for Reynolds numbers in the range $0.1 - 100$. The major aim is to study non-linear inertia effects at larger Reynolds numbers (>10) on the shear stress evolution in the fluid. This is of particular importance for blood flow as some biophysical processes in blood are sensitive to shear stresses, e.g., the initialization of blood clotting. We employ the lattice Boltzmann method for the simulations. The conclusion of the predictions is that the peak value of shear stress in the constriction grows disproportionally fast with the Reynolds number which leads to a non-linear shear stress accumulation. As a consequence, the combination of constricted blood vessel geometries and large Reynolds numbers may increase the risk of undesired blood clotting.
\end{abstract}

\maketitle


\section{Introduction}
\label{sec:introduction}

There is growing evidence that blood clotting is a dynamic process in which fluid shear stress plays an important role \cite{schneider_shear-induced_2007}. The protein von Willebrand factor (VWF) shows a conformation change when the ambient shear rate reaches values of about $5000\, \text{s}^{-1}$. It is also known that arterial plaque and stents favor the emergence of blood clots \cite{libby_current_2001, jeremias_stent_2004}. Besides biochemical reasons, one possible physical cause for clotting may be the detrimental influence of large local shear stresses in the vicinity of constrictions. Physically, the flow boundary conditions are modified by the presence of obstacles. Depending on the Reynolds number, obstacles can have a significant impact on the flow properties, even beyond the location of the obstacle. A prominent example is the Kármán vortex street which is a repeating pattern of swirling vortexes occurring in the laminar flow regime behind a bluff obstacle.

The motivation for this article is to study inertia effects on the shear stress in fluid flows perturbed by a simple obstacle and their basic implications for hemorheology. In our present investigation, we numerically model and experimentally measure the flow of a Newtonian fluid through a duct with a simple constriction at different Reynolds numbers in the laminar regime ($0.1 - 100$). We emphasize that the direct comparison of experiments and simulations is essential. Although the accuracy and applicability of the employed lattice Boltzmann method (LBM) has been proven various times \cite{chen_lattice_1998, succi_lattice_2001, sukop_lattice_2005, dnweg_lattice_2009}, the numerical results should be supported and verified by experiments.

One of the simplest symmetric, yet non-trivial, flow geometries is a duct with a bottleneck-like constriction as sketched in Fig.\ \ref{fig:geometry}. Using this obstacle, we study fundamental properties of the fluid flow at different Reynolds numbers between $0.1$ and $100$. The particular design of the constriction has been chosen for convenience since this geometry is easily produced by milling techniques (shaping solid materials with a cutter) for the experiments.

Since we are interested in the basic physical effects of inertia in a constricted geometry, we simplify the problem as much as possible. In particular, the fluid flow is assumed to be steady, and we use a Newtonian fluid (water in the experiments). For this reason, the Womersley number (a dimensionless measure for the period of pulsatile flows related to the viscous time scale) is zero, and the Reynolds number is the only relevant physical parameter. The assumption of a Newtonian fluid limits the validity of the simulations and experiments to larger blood vessels because at those scales the individual motion of the red blood cells can be neglected and the viscosity is virtually independent of the shear rate \cite{chien_shear_1970}. However, this is no severe restriction since we are mainly interested in the blood flow in human coronary arteries with average diameters of $3$--$4\, \text{mm}$ \cite{hort_size_1982} which is $1000$ times the radius of a red blood cell. Typical Reynolds numbers in coronary arteries are of order $100$ \cite{hiroyuki_hikita_low_2009}.

Due to the non-linear character of the Navier-Stokes equations, at large Reynolds numbers, abrupt changes in cross-section may lead to spatial variations of velocity and shear stress which cannot be fully understood from dimensional considerations or the Stokes equations. For this reason, we study the impact of a bottleneck-like constriction on the local properties of the fluid as a function of the Reynolds number. We are particularly interested in the spatial asymmetry and the magnitude of the shear stress. In order to emphasize the nonlinear effects arising at high Reynolds numbers, we vary the Reynolds number over three order of magnitude, $(0.1-100)$, thus covering both the fully viscous and inertial regimes. At large Reynolds numbers, the spatial flow velocity and shear stress fields are asymmetric, even in a symmetric geometry. The cause is the convective term in the Navier-Stokes equations. This asymmetry introduces a distinction of the pre- and post-constriction regions. Moreover, one can observe that the peak values of the shear stress close to the constriction increase faster than linearly with the Reynolds number, showing the significance of the inertia effects.

The article is organized as follows. The basic hydrodynamic concepts are presented in Sec.\ \ref{sec:theory}, followed by a detailled description of the experimental and numerical setup in Sec.\ \ref{sec:setup}. The observations and results are presented and discussed in Sec.\ \ref{sec:results}. Finally, the conclusions are pointed out in Sec. \ref{sec:conclusions}.


\section{Theory}
\label{sec:theory}

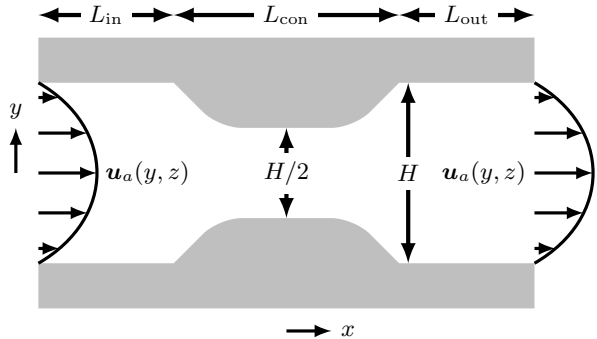
\begin{figure}
\centering
\begin{tikzpicture}[xscale=0.6,yscale=0.6]
\fill[lightgray] (0,0) -- (11,0) -- (11,1) -- (8,1)[rounded corners=0.3cm] -- (7,2) -- (4,2)[sharp corners] -- (3,1) -- (0,1) -- cycle;
\fill[lightgray,yshift=6cm,yscale=-1] (0,0) -- (11,0) -- (11,1) -- (8,1)[rounded corners=0.3cm] -- (7,2) -- (4,2)[sharp corners] -- (3,1) -- (0,1) -- cycle;
\draw[latex-latex, ultra thick] (0,6.5) -- node[midway,fill=white]{$L_{\text{in}}$} (3,6.5);
\draw[latex-latex, ultra thick] (3,6.5) -- node[midway,fill=white]{$L_{\text{con}}$} (8,6.5);
\draw[latex-latex, ultra thick] (8,6.5) -- node[midway,fill=white]{$L_{\text{out}}$} (11,6.5);
\draw[latex-latex, ultra thick] (8.2,1) -- node[midway,fill=white]{$H$} (8.2,5);
\draw[latex-latex, ultra thick] (5.5,2) -- node[midway,fill=white]{$H / 2$} (5.5,4);
\draw[very thick] (0,1) to[in=-30,out=30,distance=2cm] coordinate[pos=0.1](a) coordinate[pos=0.3](b) coordinate[pos=0.5](c) coordinate[pos=0.7](d) coordinate[pos=0.9](e) (0,5);
\draw[latex-, very thick] (a) -- (a -| 0,0);
\draw[latex-, very thick] (b) -- (b -| 0,0);
\draw[latex-, very thick] (c) -- (c -| 0,0);
\draw[latex-, very thick] (d) -- (d -| 0,0);
\draw[latex-, very thick] (e) -- (e -| 0,0);
\node[right] at (c) {$\bm u_a(y,z)$};
\draw[very thick] (11,1) to[in=-30,out=30,distance=2cm] coordinate[pos=0.1](aa) coordinate[pos=0.3](bb) coordinate[pos=0.5](cc) coordinate[pos=0.7](dd) coordinate[pos=0.9](ee) (11,5);
\draw[latex-, very thick] (aa) -- (aa -| 11,0);
\draw[latex-, very thick] (bb) -- (bb -| 11,0);
\draw[latex-, very thick] (cc) -- (cc -| 11,0);
\draw[latex-, very thick] (dd) -- (dd -| 11,0);
\draw[latex-, very thick] (ee) -- (ee -| 11,0);
\node[left] at (cc -| 11,0) {$\bm u_a(y,z)$};
\draw[-latex, very thick] (5.5,-0.5) -- (6.5,-0.5) node[right]{$x$};
\draw[-latex, very thick] (-0.5,3) -- (-0.5,4) node[above]{$y$};
\end{tikzpicture}
\caption{A 2D projection of the constriction geometry used both in the experiments and simulations. The fluid enters the geometry from the left. Numerically, the inlet and outlet velocities $\bm u_a(y, z)$ are taken from the analytic solution of the steady duct flow problem. The origin of the coordinate system is at the center of the constriction. The initial width (along $y$-axis) and height (along $z$-axis) are $H$, and the constricted width is $H/2$. The total length of the constriction is $L_{\text{con}} = 5H / 4$, and the initial inclination of the constriction walls is $45^\circ$. The rounded corners with radius $r = H/4$ are due to the milling technique used for fabricating the duct. The numerical values of the inlet and outlet duct lengths $L_{\text{in}}$ and $L_{\text{out}}$ are chosen in such a way that the flow can fully develop.}
\label{fig:geometry}
\end{figure}

The full incompressible Navier-Stokes equations in the absence of a body force density read
\be
\label{eq:navierstokes}
\rho \left(\f{\partial \bm u}{\partial t} + \bm u \cdot \nabla \bm u\right) = - \nabla p + \eta \Delta \bm u
\ee
where $\bm u$ denotes the velocity, $\rho$ the density, $p$ the pressure, and $\eta$ the viscosity of the fluid. Introducing the deviatoric shear stress tensor with components
\be
\label{eq:shearstress}
\sigma_{\alpha \beta} = \eta \left(\partial_\alpha u_\beta + \partial_\beta u_\alpha\right),
\ee
the viscous term in Eq.\ (\ref{eq:navierstokes}) can be written in the form $\eta \Delta \bm u = \nabla \cdot \bm \sigma$ for an incompressible fluid. The deviatoric shear stress tensor $\bm \sigma$ (from now on only called shear stress) is part of the total momentum flux tensor of the fluid,
\be
\label{eq:momentumflux}
M_{\alpha \beta} = p\, \delta_{\alpha \beta} + \rho\, u_\alpha u_\beta - \sigma_{\alpha \beta}.
\ee
Eq.\ (\ref{eq:navierstokes}) can then be written in the compact form $\rho\, \partial_t u_\alpha = - \partial_\beta M_{\alpha \beta}$. The first term on the right-hand-side of Eq.\ (\ref{eq:momentumflux}) is the isotropic pressure contribution, the second term denotes momentum transport due to convection (mass transport) which is only important at large Reynolds numbers.

The shear stress $\bm \sigma$ describes the momentum diffusion related to the viscosity of the fluid which is the central quantity when it comes to mechanically triggered blood clotting \cite{schneider_shear-induced_2007}. It is a second order tensor with six independent components (in this case only five because it is traceless due to the incompressibility of the fluid). Since the tensor is symmetric, it has always three real eigenvalues $\sigma_3 \leq \sigma_2 \leq \sigma_1$. For an incompressible fluid, the eigenvalues obey $\text{Tr}\, \bm \sigma = \sigma_1 + \sigma_2 + \sigma_3 = 0$. In the literature about applications of sheared fluids, usually an equivalent shear stress scalar $\sigma_{\text{eff}}$ is provided, and the tensor properties are lost. Assuming that the dynamics of VWF is not sensitive to the tensor components $\sigma_{\alpha \beta}$ but only to an effective scalar value, it arises the need for an appropriate definition. There are different possibilities to construct an effective scalar measure from the full tensor. The von Mises stress is
\be
\label{eq:vonmisesstress}
\begin{aligned}
\sigma_{\text{vM}} := \sqrt{\f 12 \sum_{\alpha, \beta} \sigma_{\alpha \beta} \sigma_{\alpha \beta}} &= \sqrt{\f 12 \left(\sigma_1^2 + \sigma_2^2 + \sigma_3^2\right)} \\
&= \sqrt{\sigma_1^2 + \sigma_3^2 + \sigma_1 \sigma_3}.
\end{aligned}
\ee
The last identity in Eq.\ (\ref{eq:vonmisesstress}) is valid if the fluid is incompressible, $\text{Tr}\, \bm \sigma = 0$. The von Mises stress plays an important role in the analysis of materials mechanics in material science, but it can also be applied to fluids. Given the shear stress tensor at a specific point, one can also ask in which direction $\bm n$ the maximum shear component $\sigma_{\text{max}}$ of the tensor can be found. From a Mohr analysis one finds
\be
\label{eq:mohrstress}
\sigma_{\text{max}} := \f{\sigma_1 - \sigma_3}{2}.
\ee
The definitions in Eqs.\ (\ref{eq:vonmisesstress}) and (\ref{eq:mohrstress}) and further comments can be found in monographs about elasticity, e.g., in \cite{sadd2009elasticity}. For an incompressible fluid, in general $\sigma_{\text{vM}} \geq \sigma_{\text{max}}$ holds, but one can show that if the eigenvalue $\sigma_2$ vanishes, $\sigma_{\text{vM}} = \sigma_{\text{max}} = \sigma_1$ since then $\sigma_3 = -\sigma_1$. One can further show that the maximum deviation between both shear stress scalars is about $15\%$. For this reason, we drop a separate discussion of $\sigma_{\text{vM}}$ and $\sigma_{\text{max}}$ and restrict ourselves to the von Mises stress. As a compact notation for the shear stress magnitude, we define
\be
\label{eq:norm_s}
\sigma = \left \lVert \bm \sigma \right \rVert := \sigma_{\text{vM}},
\ee
cf.\ Eq.\ (\ref{eq:vonmisesstress}). We emphasize that the concept of shear stress is applicable to material sciences and hydrodynamics, i.e., to solids and fluids. The definition for the velocity magnitude is the usual one,
\be
\label{eq:norm_u}
u = \left \lVert \bm u \right \rVert := \sqrt{\sum_\alpha u_\alpha u_\alpha}.
\ee

In the present article, the duct Reynolds number is defined as
\be
\label{eq:reynoldsnumber}
\text{Re} = \f{\rho H \bar u}{\eta}.
\ee
$H$ is the height and the width of the square duct, cf.\ Fig.\ \ref{fig:geometry}. The velocity scale $\bar u$ will be formally introduced in Eq.\ (\ref{eq:scalevelocity}). It is the average velocity on the inlet cross-section with area $A = H^2$ and can easily be obtained in experiments by $\dot V = A \bar u$ if the volume flux $\dot V$ of the fluid is known.

In the limit of small Reynolds numbers and a stationary situation, the left-hand-side of Eq.\ (\ref{eq:navierstokes}) is negligible, and one can write the Stokes equation
\be
\label{eq:stokes}
0 = - \nabla p + \eta \Delta \bm u.
\ee
Note that Eq.\ (\ref{eq:stokes}) is invariant under the transformation $(\bm u \to -\bm u, \nabla p \to -\nabla p)$, whereas Eq.\ (\ref{eq:navierstokes}) is not. As a consequence, the stationary velocity field for $\text{Re} = 0$ looks the same (up to its sign), when the flow of the fluid through a fixed geometry is reversed (also the signs of possible velocity boundary conditions have to be reversed). In case of a point-symmetric geometry (invariant under the transformation $\bm x \to -\bm x$), as we use it here, cf.\ Fig.\ \ref{fig:geometry}, it is easy to see that $\bm u(\bm x) = \bm u(-\bm x)$ must hold if $\text{Re} = 0$. The shear stress obeys $\bm \sigma(\bm x) = -\bm \sigma(-\bm x)$ since the spatial derivative enters its definition, Eq.\ (\ref{eq:shearstress}), and leads to an additional minus sign.  For a finite Reynolds number, the presence of the convective term $\bm u \cdot \nabla \bm u$ breaks the symmetry, and the direction of the flow can be recognized by its inertia effects and the shape of the streamlines. The Kármán vortex street is a demonstrative example for this statement: The vortexes appear only downstream.

\begin{table}
\begin{ruledtabular}
\begin{tabular}{D{.}{.}{-1}D{.}{.}{-1}D{.}{.}{-1}}
\multicolumn{1}{c}{$\text{Re}$} & \multicolumn{1}{c}{$\bar u\, [\text{mm}\, \text{s}^{-1}]$} & \multicolumn{1}{c}{$\dot V\, [\text{ml}\, \text{min}^{-1}]$} \\ \hline
0.1 & 0.05 & 0.012 \\
1   & 0.5  & 0.12 \\
10  & 5    & 1.2 \\
100 & 50   & 12 \\
\end{tabular}
\end{ruledtabular}
\caption{\label{tab:volumeflux}Experimental volumetric flow rates $\dot V$ for achieving the desired Reynolds numbers and the average inlet velocities $\bar u$. The length scale is $H = 2\, \text{mm}$, and the fluid properties are $\rho = 1000\, \text{kg}\, \text{m}^{-3}$ and $\eta = 10^{-3}\, \text{Pa}\, \text{s}$.}
\end{table}

\begin{table}
\begin{ruledtabular}
\begin{tabular}{D{.}{.}{-1}D{.}{.}{-1}D{.}{.}{-1}D{.}{.}{-1}c}
\multicolumn{1}{c}{$\text{Re}$} & \multicolumn{1}{c}{$\tau$} & \multicolumn{1}{c}{$\hat u$} & \multicolumn{1}{c}{$L_D / H$} & $L_{\text{in}}, L_{\text{out}}$ \\ \hline
0.1 & 0.9  & 0.000280 & 0.6 & $125$ \\
1   & 0.9  & 0.00280  & 0.6 & $125$ \\
10  & 0.8  & 0.0210   & 0.9 & $200$ \\
20  & 0.7  & 0.0280   & 1.4 & $200$ \\
45  & 0.6  & 0.0314   & 2.7 & $300$ \\
100 & 0.54 & 0.0280   & 5.8 & $600$ \\
\end{tabular}
\end{ruledtabular}
\caption{\label{tab:simulationparameters} Relevant simulation parameters: The diameter of the unconstricted duct in all the simulations is $H = 100$ lattice nodes. $\hat u$ is the lattice velocity at the center of the inlet and outlet cross-sections. It can be shown that for the current geometry the average velocity is $\bar u \approx 0.48 \hat u$ \cite{kruger_shear_2009}. $\tau$ is the relaxation time used for the BGK lattice Boltzmann method, and $L_D$ is the development length for a given Reynolds number according to Eq.\ (\ref{eq:developmentlength}) with $D = H$. $L_{\text{in}}$ and $L_{\text{out}}$ are the numerical values for the duct inlet and outlet lengths.}
\end{table}

In this article, we study the asymmetry introduced by finite inertia as a function of the Reynolds number. Additionally, the effects of inertia on the shear stress distribution is analyzed. For this reason, we introduce an index of distortion for the velocity and the shear stress. Knowing the analytic solutions $\bm u_a(y, z)$ and $\bm \sigma_a(y, z)$ for a fully developed flow in the duct \cite{haberman_applied_2004, latt_straight_2008, kruger_shear_2009} and taking the actual velocity $\bm u_d(y, z)$ and shear stress $\bm \sigma_d(y, z)$ at a given cross-section at axial distance $d$ from the constriction, we define
\begin{align}
\label{eq:indexvelocity}
I_u(d) &= \f{1}{A \bar u} \sum_{y,z} \left\lVert \bm u_a(y,z) - \bm u_d(y,z) \right\lVert, \\
\label{eq:indexstress}
I_\sigma(d) &= \f{1}{A \bar \sigma} \sum_{y,z} \left\lVert \bm \sigma_a(y,z) - \bm \sigma_d(y,z) \right\lVert
\end{align}
where the norm $\lVert\cdot\lVert$ for the shear stress and the velocity has been defined in Eqs.\ (\ref{eq:norm_s}) and (\ref{eq:norm_u}). The quantities
\begin{align}
\label{eq:scalevelocity}
\bar u &= \f 1A \sum_{y,z} \left\lVert \bm u_a(y,z) \right\lVert, \\
\label{eq:scalestress}
\bar \sigma &= \f 1A \sum_{y,z} \left\lVert\bm \sigma_a(y,z) \right\lVert
\end{align}
are the velocity and shear stress scales, defined as the averages on the inlet cross-section. They obey $\bar u \propto \text{Re}$ and $\bar \sigma \propto \text{Re}$ and are control quantities which the numerical results will be related to.

In Stokes flow, $\text{Re} = 0$, the fluid velocity and shear stress distributions look the same before and behind the constriction (up to signs). For this reason, Eqs.\ (\ref{eq:indexvelocity}) and (\ref{eq:indexstress}) are invariant if the coordinate system is transformed according to $\bm x \to -\bm x$. Thus, the indexes of distortion $I_u$ and $I_\sigma$ do not change when the coordinate system is transformed. At finite Reynolds number, however, the symmetry is broken, and inlet and outlet flow profiles differ. Thus, we expect that $I_u(d) \not= I_u(-d)$ and $I_\sigma(d) \not= I_\sigma(-d)$ when $\text{Re} \not= 0$. As we will show in Section \ref{sec:results}, the slopes of $I_u(d)$ and $I_\sigma(d)$ are well captured by a simple decaying exponential. For this reason, it is natural to define a range of decay $\lambda$ for each exponential. It is given by the distance from the constriction after which the exponential decays to $\exp(-1)$ of its initial value. In order to distinguish the ranges of decay for the velocity and the shear stress, we denote both quantities $\lambda_u$ and $\lambda_\sigma$, respectively. This way, the indexes of distortion can be approximated by
\begin{align}
\label{eq:exponential_u}
I_u(d) &= I_u(0) \exp(-d / \lambda_u), \\
\label{eq:exponential_s}
I_\sigma(d) &= I_\sigma(0) \exp(-d / \lambda_\sigma).
\end{align}

Related to the definition of $\lambda_u$ is the problem of the flow development length $L_D$ in a 2D channel or a 3D pipe. It has been thoroughly discussed in the literature analytically, numerically, and experimentally due to its important implications in engineering \cite{friedmann_laminar_1968, atkinson_low_1969, durst_development_2005}. The common approach is to impose a constant velocity profile at the inlet of a pipe with circular cross-section and diameter $D$ and find the axial distance $L_D$ from the inlet at which the central velocity has reached $99\%$ of its fully developed value. The parameter $L_D / D$ is a function of the Reynolds number. \citet{durst_development_2005} have proposed the relation
\be
\label{eq:developmentlength}
\f{L_D}{D} = \left(0.619^{1.6} + (0.0567\, \text{Re})^{1.6}\right)^{1/1.6}
\ee
for a Newtonian fluid in a pipe and $\text{Re} = \rho D \bar u / \eta$. This equation is valid for all Reynolds numbers as long as the flow is laminar, and the numerical error is reported to be $<3\%$. In the present simulations, there is a slightly different situation: The geometry is a duct with quadratic cross-section, and the velocity profile at the constriction is not constant. However, it has turned out that Eq.\ (\ref{eq:developmentlength}) is also a good approximation for the presented problem. The development lengths obtained from Eq.\ (\ref{eq:developmentlength}) and using $H$ instead of $D$ are shown in Tab.\ \ref{tab:simulationparameters}. Those values act as a guideline for the experiments and simulations to assure that the duct before and behind the constriction is sufficiently long.


\section{Experimental and numerical setup}
\label{sec:setup}

\begin{figure}
\begin{tikzpicture}[xscale=1.4,yscale=1.4]
\small
\fill[lightgray] (-2.5,-1.5) -- (2.5,-1.5) -- (2.5,-1) -- (1.25,-1)[rounded corners=0.35cm] -- (0.8,-0.5) -- (-0.8,-0.5)[sharp corners] -- (-1.25,-1) -- (-2.5,-1) -- cycle;
\fill[lightgray] (-2.5,1.5) -- (2.5,1.5) -- (2.5,1) -- (1.25,1)[rounded corners=0.35cm] -- (0.8,0.5) -- (-0.8,0.5)[sharp corners] -- (-1.25,1) -- (-2.5,1) -- cycle;
\draw[dashed] (-2.5,0) -- (2.5,0) node[right]{$0.0$};
\draw[dashed] (-2.5,0.2) -- (2.5,0.2) node[right]{$0.2$};
\draw[dashed] (-2.5,0.4) -- (2.5,0.4) node[right]{$0.4$};
\draw[dashed] (-2.5,0.6) -- (2.5,0.6) node[right]{$0.6$};
\draw[dashed] (-2.5,0.8) -- (2.5,0.8) node[right]{$0.8$};
\draw[dotted,gray] (-2.25,-1.5) -- (-2.25,1.85);
\draw[dashed] (-1.85,-1.5) -- (-1.85,1.5) node[above]{$-1.85$};
\draw[dotted,gray] (-1.45,-1.5) -- (-1.45,1.85);
\draw[dotted,gray] (-0.35,-1.5) -- (-0.35,1.85);
\draw[dashed] (0.05,-1.5) -- (0.05,1.5) node[above]{$0.05$};
\draw[dotted,gray] (0.45,-1.5) -- (0.45,1.85);
\draw[dotted,gray] (1.45,-1.5) -- (1.45,1.85);
\draw[dashed] (1.85,-1.5) -- (1.85,1.5) node[above]{$1.85$};
\draw[dotted,gray] (2.25,-1.5) -- (2.25,1.85);
\draw[latex-latex,gray] (-2.25,1.85)--(-1.45,1.85) node[above,midway]{$0.8$};
\draw[latex-latex,gray] (-0.35,1.85)--(0.45,1.85) node[above,midway]{$0.8$};
\draw[latex-latex,gray] (1.45,1.85)--(2.25,1.85) node[above,midway]{$0.8$};
\foreach \y in {0.,0.2,...,0.8}
\draw[fill] (-1.85,\y) circle (0.03);
\foreach \y in {0.,0.2,...,0.8}
\draw[fill] (1.85,\y) circle (0.03);
\foreach \y in {0.,0.2,...,0.4}
\draw[fill] (0.05,\y) circle (0.03);
\draw[-latex, very thick] (0,-1.75) -- (0.5,-1.75) node[right]{$x$ (flow direction)};
\draw[-latex, very thick] (-2.75,0) -- (-2.75,0.5) node[above]{$y$};
\end{tikzpicture}
\caption{Locations of the velocity measurements in the experiments (black dots). All positions and distances are given in units of $\text{mm}$. The origin is located at the center of the constriction. The flow enters from the left. The velocities have been measured at three positions along the $x$-axis ($x = -1.85\, \text{mm}$, $0.05\, \text{mm}$, and $1.85\, \text{mm}$) and five positions along the $y$-axis ($y = 0$ to $0.8\, \text{mm}$ in steps of $0.2\, \text{mm}$). In the constriction, only three data points have been taken ($y = 0$, $0.2\, \text{mm}$, and $0.4\, \text{mm}$). In order to compute the velocities, the times of travel of the tracer particles between the dotted lines have been measured ($\Delta x = 0.8\, \text{mm}$ and $\Delta y = 0$, respectively). The velocity data is shown in Tab.\ \ref{tab:velocitydata}.}
\label{fig:measurements}
\end{figure}
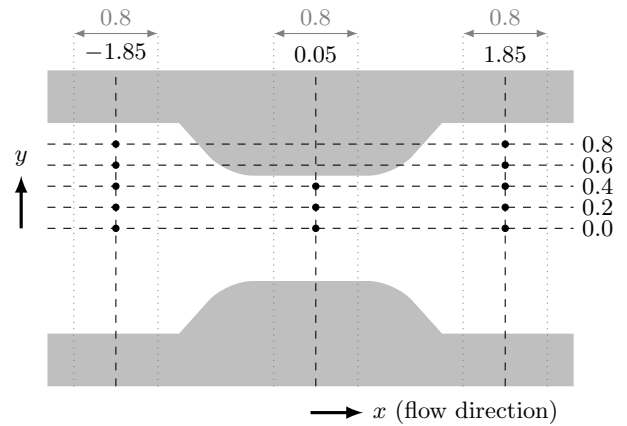

The employed geometry is a square duct (width = height = $H$) in the $yz$-plane. A 2D projection is shown in Fig.\ \ref{fig:geometry}. The flow enters at the inlet in $x$-direction. A constriction of total length $L_{\text{con}} = 5H/4$ is located halfway between the inlet and the outlet. In the constriction, the width of the duct (along the $y$-axis) is decreased, but the height (along the $z$-axis) is not changed. The constricted width is $H/2$, leading to an average flux velocity two times larger than in the main duct. Due to the milling technique employed for the experiments, the inner edges of the constriction are rounded with radius $r = H/4$. The origin of the coordinate system is always located at the center of the constriction.

\subsection{Experiments}
\label{subsec:experiments}

The experimental setup consists of the duct, cf.\ Fig.\ \ref{fig:geometry}, a syringe pump and tubes for connecting the pump to the duct. The height of the duct is $H = 2\, \text{mm}$. The flux is driven at a desired rate by application of a syringe pump (NE-1000, New Era Pump Systems, Inc., NY, USA). Connection between the pump and the duct succeeds over tubes connecting the syringe needle to the inlet. The duct itself consists of two parts. The upper part is made of polydimethylsiloxane (PDMS), casted into a mold produced by milling. This part is then converted to a completely closed duct by being attached to a microscope glass slide which plays the role of the lower deck of the duct. Inlet and outlet are punched into the duct before attachment of the PDMS to the glass slide. Attachment occurs by plasma oxidation of the PDMS and the glass slide.

The local fluid velocity in the duct is measured by tracking polystyrene beads (Polysciences, Inc., Warrington, PA, USA) with a diameter of $10\, \mu \text{m}$ which are suspended in the carrier fluid (water, density $\rho = 1000\, \text{kg}\, \text{m}^{-3}$ and viscosity $\eta = 10^{-3}\, \text{Pa}\, \text{s}$ at $20^\circ\text{C}$).

The experiments are conducted on a Zeiss Axiovert 200 inverted microscope typically using a 2.5x objective. The velocities of the beads are observed at a height of $1\, \text{mm}$ over the bottom deck of the duct (i.e., in the middle between bottom and top). For each Reynolds number, a video of the flux inside the duct is made using an ultrafast camera (Fastcam, Photron, CA, USA). The videos are analyzed by the software `Image J' afterwards.

Since the fluid properties and the spatial scale $H$ are fixed, the Reynolds number, Eq.\ (\ref{eq:reynoldsnumber}), can only be changed by choosing a mean velocity of the fluid. Four different Reynolds numbers have been investigated experimentally. The corresponding volume rates are shown in Tab.\ \ref{tab:volumeflux}.

\subsection{Simulations}
\label{subsec:simulations}

\begin{table*}
\begin{ruledtabular}
\begin{squeezetable}
\begin{tabular}{*{4}{D{.}{.}{-1}}r*{2}{D{.}{.}{-1}}r*{2}{D{.}{.}{-1}}r*{2}{D{.}{.}{-1}}r*{2}{D{.}{.}{-1}}r}
\multicolumn{2}{c}{position} &
\multicolumn{3}{c}{$\text{Re} = 0.1$} &
\multicolumn{3}{c}{$\text{Re} = 1$} &
\multicolumn{3}{c}{$\text{Re} = 10$} &
\multicolumn{3}{c}{$\text{Re} = 100$} \\
\multicolumn{1}{c}{$x[\text{mm}]$} & \multicolumn{1}{c}{$y[\text{mm}]$} &
\multicolumn{1}{c}{exp} & \multicolumn{1}{c}{sim} & \multicolumn{1}{c}{dev} &
\multicolumn{1}{c}{exp} & \multicolumn{1}{c}{sim} & \multicolumn{1}{c}{dev} &
\multicolumn{1}{c}{exp} & \multicolumn{1}{c}{sim} & \multicolumn{1}{c}{dev} &
\multicolumn{1}{c}{exp} & \multicolumn{1}{c}{sim} & \multicolumn{1}{c}{dev}\\ \hline
-1.85 & 0.0 & 0.092 & 0.109  & $16\%$ & 1.2  & 1.08  & $11\%$ & 9.8 & 10.6 & $8\%$  & 87  & 106  & $18\%$ \\
-1.85 & 0.2 & 0.088 & 0.104  & $15\%$ & 1.1  & 1.03  & $7\%$  & 9.4 & 10.2 & $8\%$  & 85  & 101  & $16\%$ \\
-1.85 & 0.4 & 0.076 & 0.0894 & $15\%$ & 0.98 & 0.891 & $10\%$ & 8.9 & 8.85 & $1\%$  & 76  & 88.7 & $14\%$ \\
-1.85 & 0.6 & 0.054 & 0.0669 & $19\%$ & 0.81 & 0.669 & $21\%$ & 6.0 & 6.69 & $10\%$ & 63  & 67.1 & $6\%$  \\
-1.85 & 0.8 & 0.038 & 0.0370 & $3\%$  & 0.50 & 0.371 & $35\%$ & 4.1 & 3.70 & $11\%$ & 43  & 35.8 & $20\%$ \\ \hline
0.05  & 0.0 & 0.17  & 0.199  & $15\%$ & 1.9  & 1.99  & $5\%$  & 23  & 19.4 & $19\%$ & 151 & 161  & $6\%$  \\
0.05  & 0.2 & 0.14  & 0.165  & $15\%$ & 1.5  & 1.65  & $9\%$  & 18  & 16.1 & $12\%$ & 144 & 150  & $4\%$  \\
0.05  & 0.4 & 0.080 & 0.0662 & $21\%$ & 0.93 & 0.662 & $40\%$ & 6.7 & 6.55 & $2\%$  & 91  & 71.1 & $28\%$ \\ \hline
1.85  & 0.0 & 0.085 & 0.109  & $22\%$ & 1.0  & 1.10  & $9\%$  & 14  & 12.8 & $9\%$  & 144 & 155  & $7\%$  \\
1.85  & 0.2 & 0.089 & 0.104  & $17\%$ & 0.92 & 1.05  & $12\%$ & 10  & 11.9 & $16\%$ & 140 & 140  & $0\%$  \\
1.85  & 0.4 & 0.071 & 0.0894 & $21\%$ & 0.91 & 0.897 & $2\%$  & 10  & 9.58 & $4\%$  & 103 & 95.2 & $8\%$  \\
1.85  & 0.6 & 0.065 & 0.0669 & $3\%$  & 0.59 & 0.669 & $12\%$ & 5.8 & 6.57 & $12\%$ & 32  & 42.8 & $25\%$ \\
1.85  & 0.8 & 0.045 & 0.0370 & $22\%$ & 0.41 & 0.369 & $11\%$ & 4.8 & 3.35 & $43\%$ & 12  & 8.46 & $42\%$ \\
\end{tabular}
\end{squeezetable}
\end{ruledtabular}
\caption{\label{tab:velocitydata} Measured and simulated velocities in the constriction at selected positions $(x,y)$ midway between the bottom and top walls, cf.\ Fig.\ \ref{fig:measurements}. All velocities are given in units of $\text{mm}\, \text{s}^{-1}$. The deviations $|u_{\text{exp}} - u_{\text{sim}}| / u_{\text{sim}}$ are also shown.}
\end{table*}

The simulations were conducted with the lattice Boltzmann method (LBM) using a D3Q19 BGK model \cite{qian_lattice_1992}. There exist excellent introductory articles \cite{he_lattice_1997, luo2000lga, ladd2001lbs}, monographs \cite{succi_lattice_2001, sukop_lattice_2005}, reviews \cite{chen_lattice_1998, raabe_topical_2004}, and various articles about applications, e.g., \cite{varnik_roughness-induced_2007, varnik_wetting_2008, kruger_submitted_efficient}.

In order to capture the physical boundary conditions of both the constriction-fluid surface (no slip) and the inlet and outlet cross-sections of the simulation box (fully developed flow), we employ the standard LBM bounce-back boundary condition \cite{ladd2001lbs} for the former and velocity boundary conditions for the latter case. At the inlet and outlet of the computational box, a fully developed velocity profile $\bm u_a(y,z)$ is imposed. The velocity boundary condition used has been proposed by \citet{latt_straight_2008}. We have chosen this approach due to its simple and straightforward implementation in three-dimensional LBM simulations. The analytic form of the stationary, fully developed flow profile for a rectangular duct is discussed in \cite{haberman_applied_2004, latt_straight_2008, kruger_shear_2009}.

We compute the pressure $p$, velocity vector $\bm u$, and the full shear stress tensor $\bm \sigma$ in the entire numerical grid. From this data, we can calculate the effects of inertia on the spatial velocity and shear stress distributions. We trace the maximum values of velocity and shear stress. Furthermore, we compute the indexes of distortion $I_u$ and $I_\sigma$, Eqs.\ (\ref{eq:indexvelocity}) and (\ref{eq:indexstress}), as function of the axial distance $d$ from the constriction, both before and behind the constriction.

In the simulations, $H$ corresponds to 100 lattice nodes to ensure a sufficiently high spatial resolution. The simulations are terminated when the relative change of velocity
\be
\f{\delta u}{\bar u} = \f{1}{N \bar u} \sqrt{\sum_{x,y,z} \left(\bm u_{\text{new}}(x,y,z) - \bm u_{\text{old}}(x,y,z)\right)^2}
\ee
becomes smaller than $10^{-10}$ between two successive time steps where $N$ is the total number of lattice nodes. This condition guarantees that the flow is stationary. In order to take into account the development length $L_D$ of a non-developed flow, Eq.\ (\ref{eq:developmentlength}), we allow the flow to relax towards inlet and outlet by extending the duct geometry correspondingly. If the inlet or outlet is too short, unphysical hydrodynamic interactions with the boundaries, such as reflections, arise. For convenience, the numerical inlet and outlet lengths are always identical, $L_{\text{in}} = L_{\text{out}}$. The rrelevantsimulation parameters are given in Tab.\ \ref{tab:simulationparameters}.


\section{Results}
\label{sec:results}

\begin{figure*}
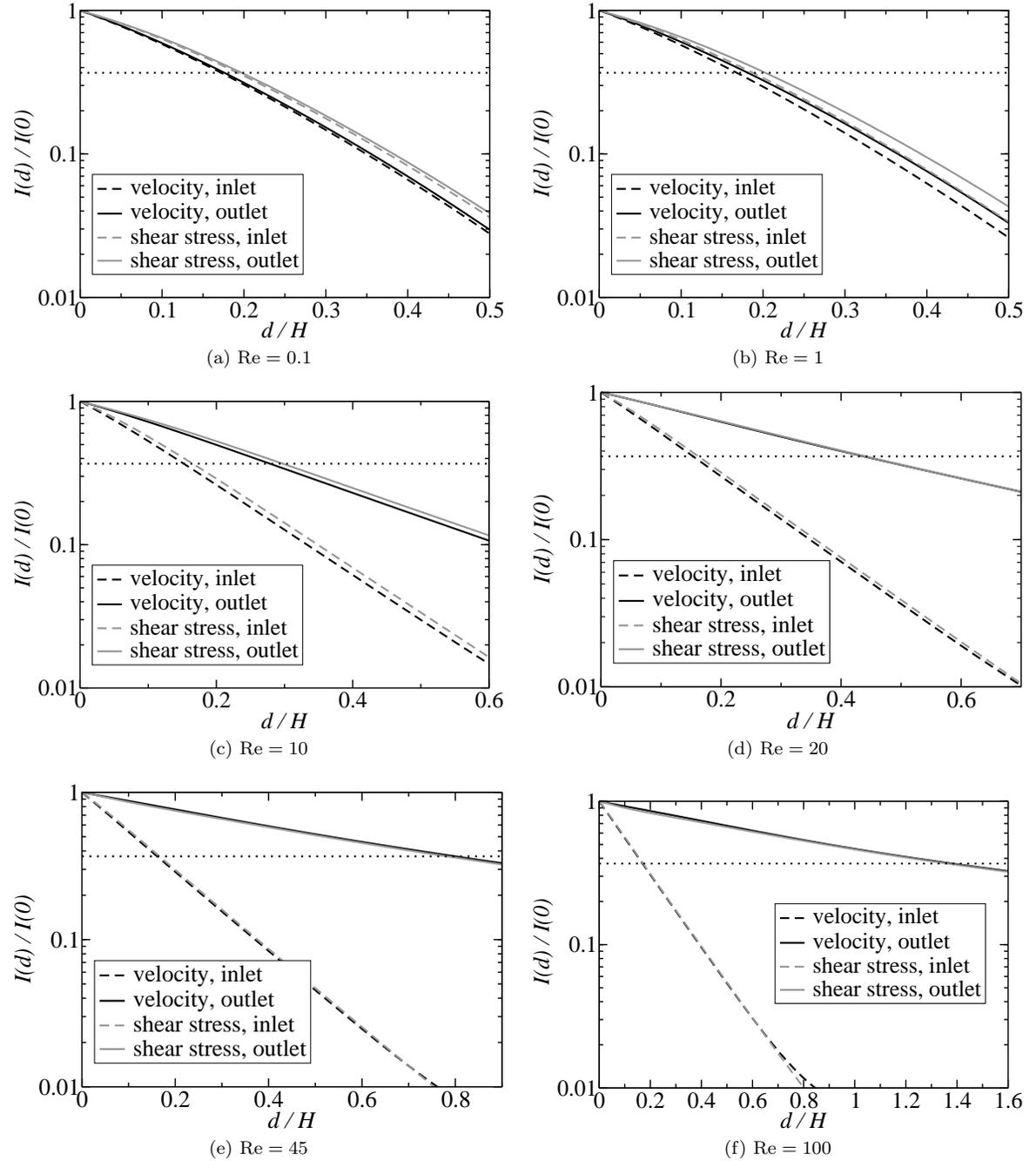

\centering
\subfigure[\label{fig:Re0.1} $\text{Re} = 0.1$]{\includegraphics[width=7.5cm,clip=true]{plots01}} \quad
\subfigure[\label{fig:Re1} $\text{Re} = 1$]{\includegraphics[width=7.5cm,clip=true]{plots1}} \\
\subfigure[\label{fig:Re10} $\text{Re} = 10$]{\includegraphics[width=7.5cm,clip=true]{plots10}} \quad
\subfigure[\label{fig:Re20} $\text{Re} = 20$]{\includegraphics[width=7.5cm,clip=true]{plots20}} \\
\subfigure[\label{fig:Re45} $\text{Re} = 45$]{\includegraphics[width=7.5cm,clip=true]{plots45}} \quad
\subfigure[\label{fig:Re100} $\text{Re} = 100$]{\includegraphics[width=7.5cm,clip=true]{plots100}}
\caption{Presentation of the normalized indexes of distortion $I_u$ and $I_\sigma$ extracted from the simulation data. The indexes of distortion are defined as the relative deviations of the velocity/shear stress profiles at a given cross-section with a distance $d$ from the constriction compared to the reference profile at the inlet, cf.\ Eqs.\ (\ref{eq:indexvelocity}) and (\ref{eq:indexstress}). For each Reynolds number, the normalized indexes of distortion are logarithmically plotted as function of the distance $d/H$ from the beginning of the constriction to the inlet and from the end of the constriction to the outlet, respectively. Additionally, the $\exp(-1)$-level is marked (horizontal, dotted line), defining the ranges $\lambda_u / H$ and $\lambda_\sigma / H$. Note that the ranges on the $d$-axis are different for the subfigures.}
\label{fig:indexdistortion}
\end{figure*}

\begin{figure}
\centering
\subfigure[\label{subfig:lines1} $\text{Re} = 1$]{\includegraphics[width=8.5cm,clip=true]{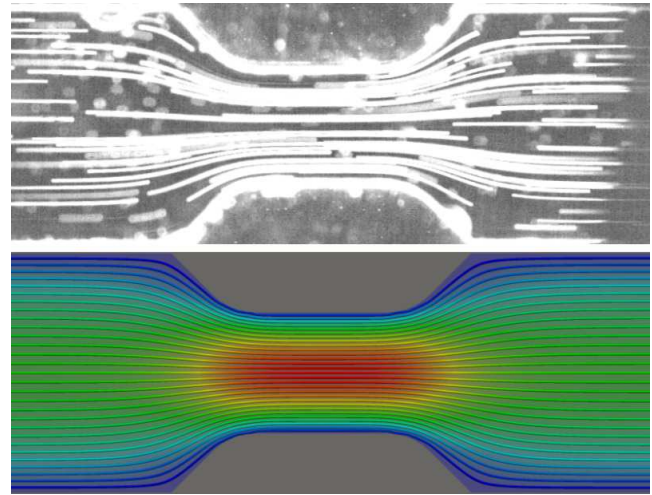}} \\
\subfigure[\label{subfig:lines100} $\text{Re} = 100$]{\includegraphics[width=8.5cm,clip=true]{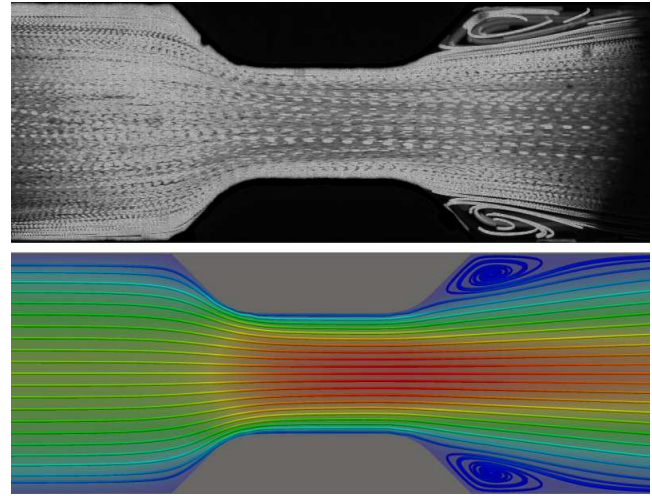}}
\caption{(Color online) The streamlines at \subref{subfig:lines1} $\text{Re} = 1$ and \subref{subfig:lines100} $\text{Re} = 100$ seen in the experiments (top) and in the simulations (bottom) at $z = 0$ (midway between bottom and top walls), respectively. The fluid enters from the left. The colors in the simulation figures correspond to the velocity magnitudes, cf.\ Fig.\ \ref{fig:streamlines_01_100}.}
\label{fig:streamlines_exp_sim}
\end{figure}

\begin{figure}
\centering
\subfigure[\label{subfig:comp_u} streamlines and velocity magnitude]{\includegraphics[width=8.5cm,clip=true]{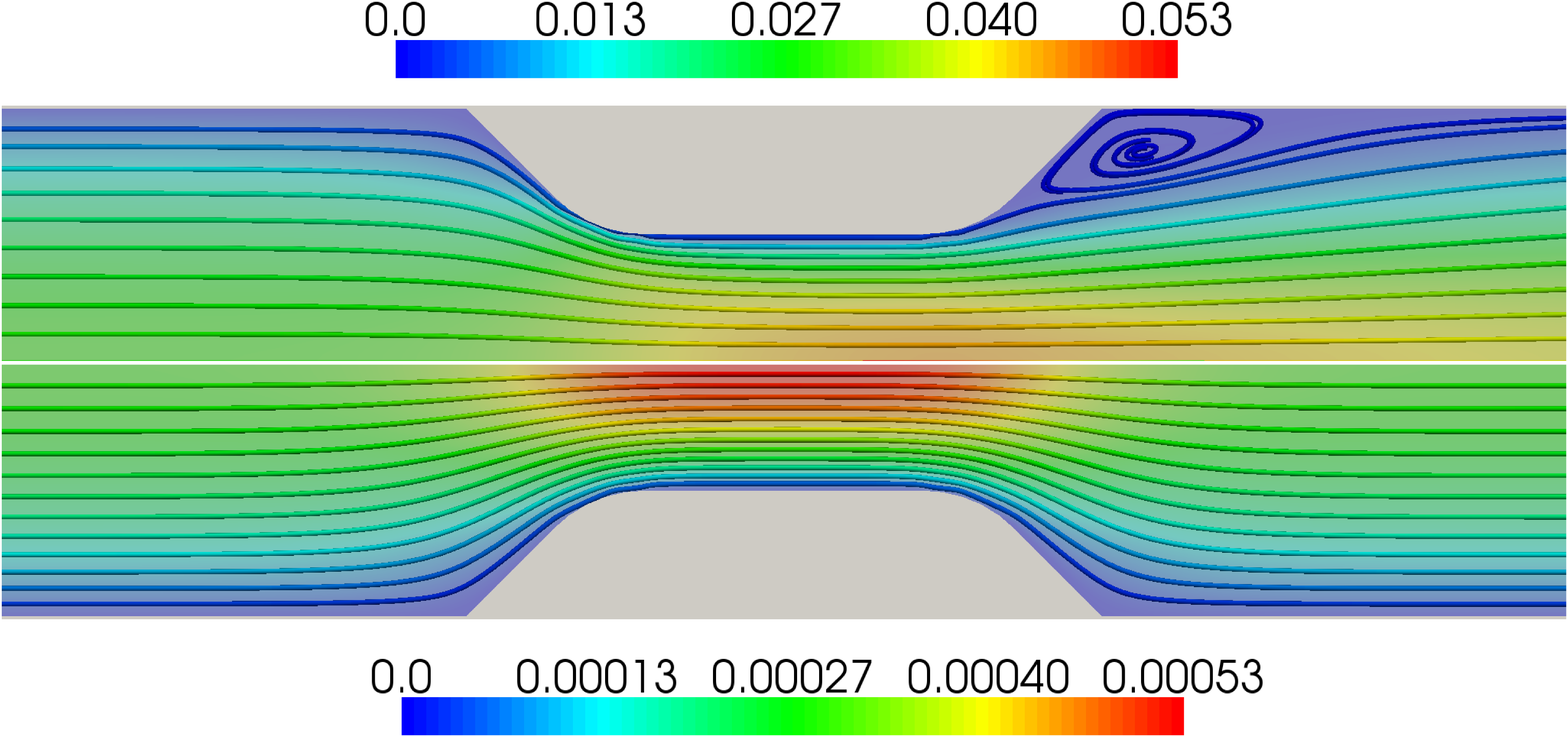}} \\
\subfigure[\label{subfig:comp_s} shear stress magnitude]{\includegraphics[width=8.5cm,clip=true]{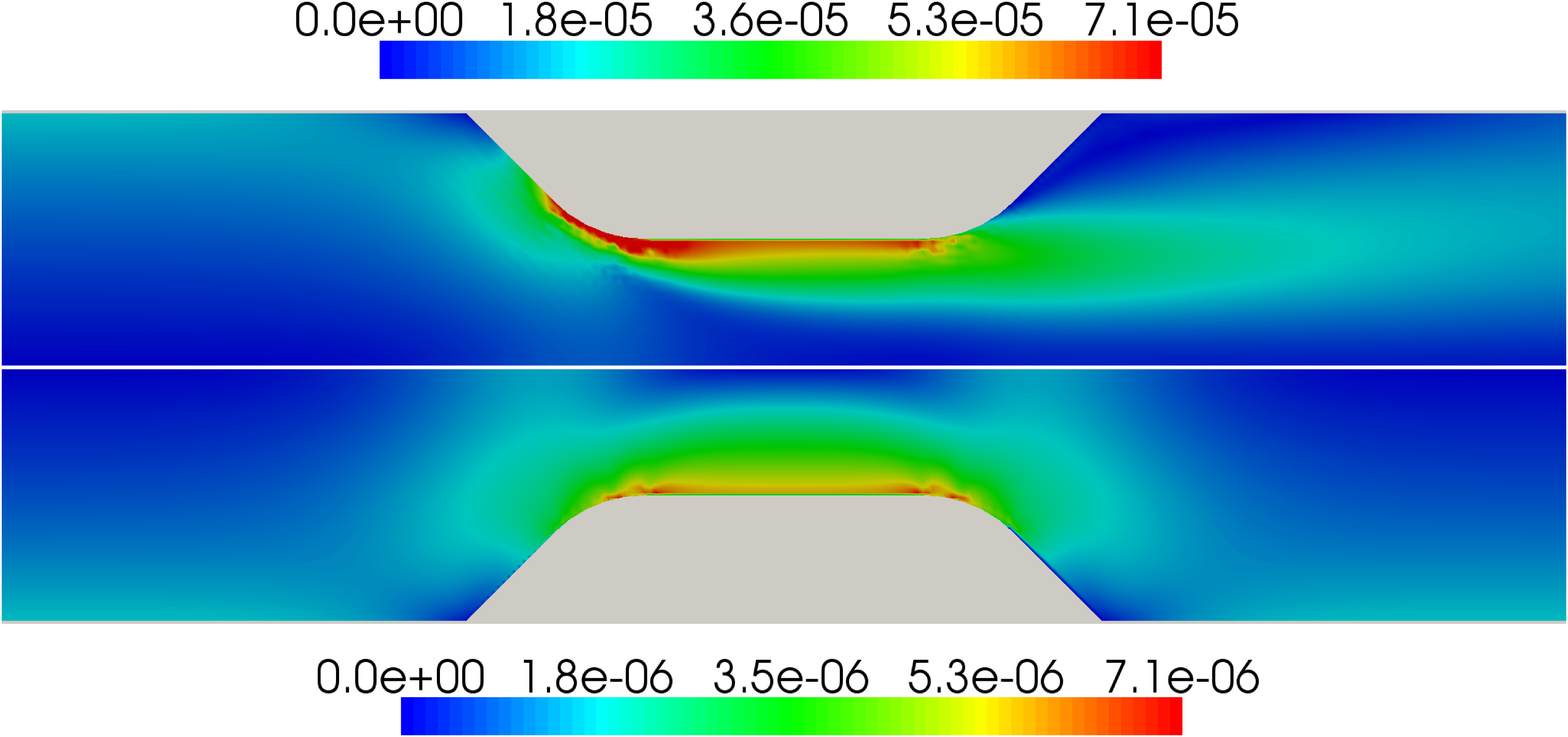}}
\caption{(Color online) Direct comparison of the simulation results: \subref{subfig:comp_u} the streamlines and velocity magnitudes and \subref{subfig:comp_s} the shear stress magnitudes at $z = 0$ (midway between bottom and top walls) for $\text{Re} = 0.1$ (bottom) and $\text{Re} = 100$ (top), respectively. The fluid enters from the left. For convenience, the colors for the magnitudes have been chosen in such a way that equal values of $u / \bar u$ or $\sigma / \bar \sigma$ have the same color. Numbers are lattice values.}
\label{fig:streamlines_01_100}
\end{figure}

\begin{table}
\begin{ruledtabular}
\begin{tabular}{D{.}{.}{-1}D{.}{.}{-1}D{.}{.}{-1}D{.}{.}{-1}D{.}{.}{-1}}
\multirow{2}*{Re} & \multicolumn{2}{c}{$\lambda_u / H$} & \multicolumn{2}{c}{$\lambda_\sigma / H$} \\
 & \multicolumn{1}{c}{inlet} & \multicolumn{1}{c}{outlet} & \multicolumn{1}{c}{inlet} & \multicolumn{1}{c}{outlet} \\ \hline
0.1 & 0.18 & 0.18 & 0.20 & 0.20 \\
1   & 0.17 & 0.19 & 0.19 & 0.21 \\
10  & 0.16 & 0.28 & 0.17 & 0.30 \\
20  & 0.16 & 0.44 & 0.17 & 0.45 \\
45  & 0.17 & 0.80 & 0.17 & 0.79 \\
100 & 0.17 & 1.38 & 0.18 & 1.37 \\
\end{tabular}
\end{ruledtabular}
\caption{\label{tab:rangeofdecay} Ranges of decay $\lambda_u / H$ and $\lambda_\sigma / H$ towards the inlet and outlet extracted from the simulation data. The ranges are defined as the distances from the constriction at which the indexes of distortion $I_u$ and $I_\sigma$ drop to $\exp(-1)$ of their value directly at the constriction. The outlet data is also illustrated in Fig.\ \ref{fig:ranges}.}
\end{table}

\begin{figure}
\centering
\includegraphics[width=8.5cm,clip=true]{rod}
\caption{The computed ranges of decay $\lambda_u$, $\lambda_\sigma$ in the outlet direction from Tab.\ \ref{tab:rangeofdecay} are shown as function of the Reynolds number. For $\text{Re} \leq 1$, the ranges are constant, and the Stokes approximation is valid. For $\text{Re} > 1$, inertia effects are important and the relaxation of the fluid is significantly delayed. The results for $\lambda_u$ and $\lambda_\sigma$ are virtually identical for a given Reynolds number.}
\label{fig:ranges}
\end{figure}

\begin{figure}
\centering
\subfigure[\label{subfig:lines_1} $y / H = 0.005$]{\includegraphics[width=7cm,clip=true]{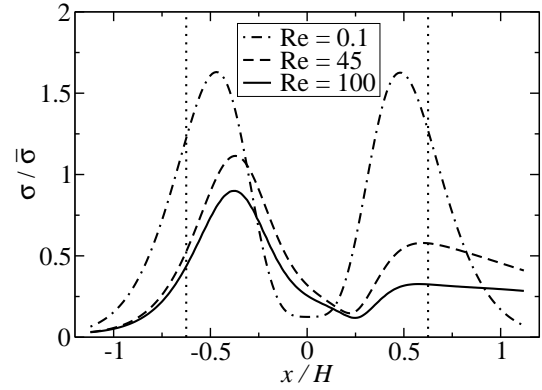}} \\
\subfigure[\label{subfig:lines_2} $y / H = 0.125$]{\includegraphics[width=7cm,clip=true]{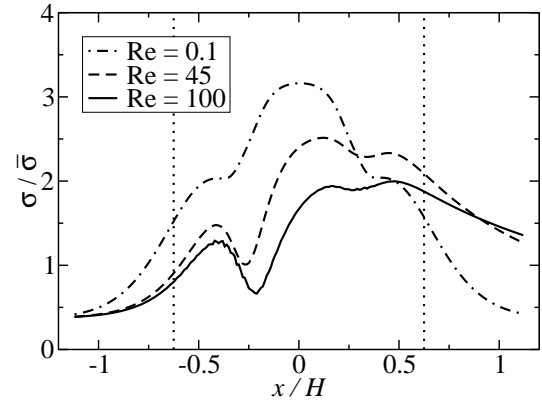}} \\
\subfigure[\label{subfig:lines_3} $y / H = 0.245$]{\includegraphics[width=7cm,clip=true]{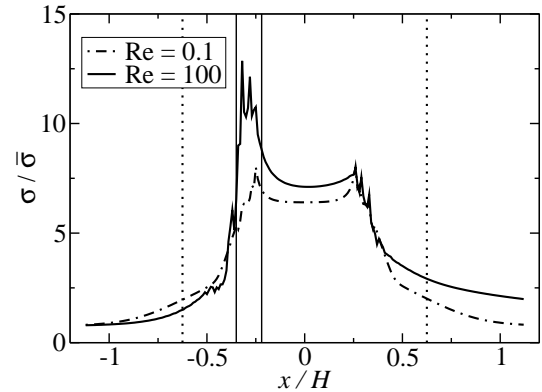}}
\caption{The shear stress $\sigma / \bar \sigma$ as a function of axial position $x$ in the constriction at $z = 0$ (midway between bottom and top walls). For the Reynolds numbers $0.1$, $45$, and $100$, the spatial shear stress evolution for three different lateral positions, \subref{subfig:lines_1} $y/H = 0.005$ (close to the central axis), \subref{subfig:lines_2} $y/H = 0.125$ (halfway between the central axis and the constricted wall), and \subref{subfig:lines_3} $y/H = 0.245$ (close to the constricted wall), is shown ($y = 0$ corresponds to the central axis and $y / H = 0.25$ to the constricted wall). The center of the constriction is located at $x = 0$. The vertical dotted lines mark the beginning and end of the constriction. For convenience, only $\text{Re} = 0.1$ and $100$ are shown in \subref{subfig:lines_3}. The solid vertical lines mark the averaging interval in order to obtain $\sigma_{\text{max}} / \bar \sigma$, Eq.\ (\ref{eq:peak_stress}). For comparison, see also Fig.\ \ref{subfig:comp_s}.}
\label{fig:stresslines}
\end{figure}

\begin{figure}
\centering
\subfigure[\label{subfig:acc_u} maximum velocity magnitudes]{\includegraphics[width=8.5cm,clip=true]{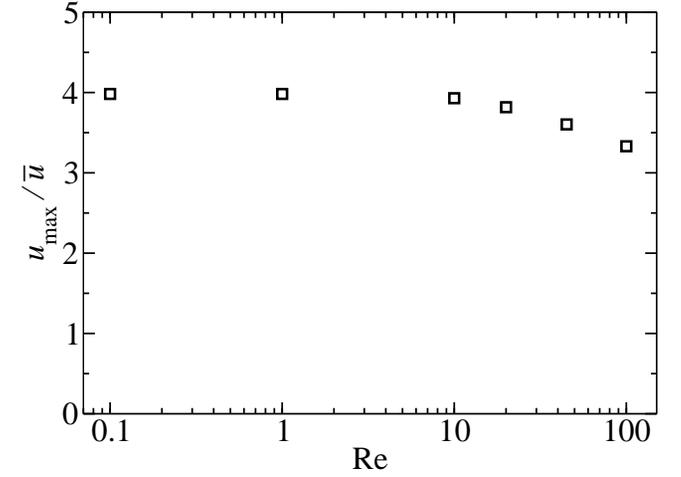}} \\
\subfigure[\label{subfig:acc_s} maximum shear stress magnitudes]{\includegraphics[width=8.5cm,clip=true]{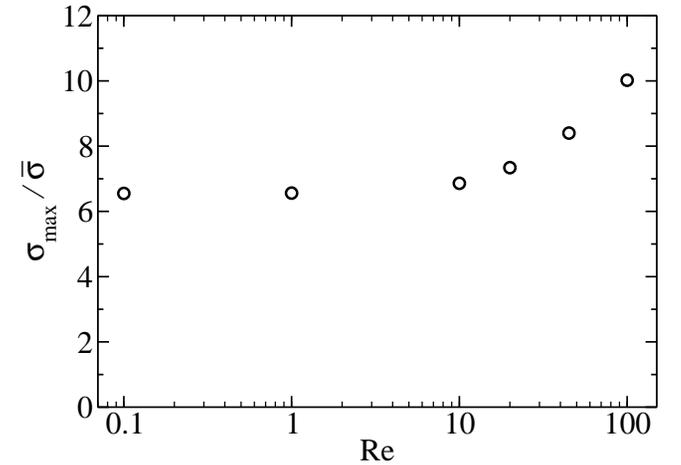}}
\caption{Maximum magnitudes of \subref{subfig:acc_u} the velocity and \subref{subfig:acc_s} the shear stress, normalized by the scales $\bar u$ and $\bar \sigma$ defined on the inlet, Eqs.\ (\ref{eq:scalevelocity}) and (\ref{eq:scalestress}). At small Reynolds numbers, both quantities do not depend on $\text{Re}$, and inertia is negligible. At larger $\text{Re}$, the shear stress increases while the velocity drops. Those are effects caused by the non-linear terms in the Navier-Stokes equations.}
\label{fig:accumulation}
\end{figure}

\paragraph{Experimental results and comparison to simulations}

In Fig.\ \ref{fig:measurements}, the locations of the velocity measurements in the experiments are presented. The velocities have been estimated by measuring the time of travel of representative tracer particles (cf.\ Sec.\ \ref{subsec:experiments}) between two positions along the $x$-axis ($\Delta x \approx 0.8\, \text{mm}$). The resultant velocity is assumed to be the velocity midway between the two points. The experimental data is shown together with the corresponding velocities from the simulations and the relative deviations $|u_{\text{exp}} - u_{\text{sim}}| / u_{\text{sim}}$ in Tab.\ \ref{tab:velocitydata}. Some streamlines found in the experiments and simulations are visualized in Figs.\ \ref{fig:streamlines_exp_sim} and \ref{fig:streamlines_01_100}. The experimental and simulated results at Reynolds numbers $1$ and $100$ are shown in Fig.\ \ref{fig:streamlines_exp_sim}. In Fig.\ \ref{fig:streamlines_01_100}, the simulated flow fields for the Reynolds numbers $0.1$ and $100$ are directly compared.

The experiments have been carried out with great care. However, as can be seen from Tab.\ \ref{tab:velocitydata}, the quantitative comparison of the experimental and simulation velocity data reveals some deviations. The major reason is that the velocities cannot be measured locally in the experiments. Instead, the motion of the tracer beads is followed over a finite distance of $0.4 H$ ($0.8\, \text{mm}$), and the velocity at the middle of this line is assumed to be the average velocity, cf.\ Fig.\ \ref{fig:measurements}. This approach can only be accurate if the length over which the particles are observed is small compared to the typical length for the change of the velocity field. This characteristic length is of order $H/4$ which is half the width of the duct inside the constriction. Consequently, there is an intrinsic uncertainty in the velocity measurement. The particles do not always move on straight lines which can be recognized from the shape of the streamlines, cf.\ Fig.\ \ref{fig:streamlines_exp_sim}. This makes it hard to achieve a good estimate for the local velocities even when the time resolution of the measurements is high. Especially close to the walls ($y = 0.8\, \text{mm}$ before and behind and $y = 0.4\, \text{mm}$ inside the constriction), the deviations are expected to be larger. The reason is that the velocity gradient is maximum in the vicinity of the walls. If the position of the tracer beads is slightly shifted along the $y$-axis, this will lead to a large uncertainty in the velocity measurement. This trend can clearly be recognized in Tab.\ \ref{tab:velocitydata}. An additional, yet minor, reason for the deviations is that the tracer particles do not necessarily move exactly in the plane midway between the bottom and top walls ($z = 0$). However, it is encouraging to see that the qualitative shape of the experimental streamlines is recovered by the computer simulations. Especially the shape of the vortexes at $\text{Re} = 100$, cf.\ Fig.\ \ref{subfig:lines100}, is correctly reproduced. Taking those considerations into account, the agreement between experiments and simulations is satisfactory.

In order to test the confidence in the simulations, we have decreased the numerical resolution from $H = 100$ to $H = 80$ and $40$ (data not shown). We observe that the numerical results for $H = 80$ are virtually identical to those for $H = 100$ indicating that the resolution is sufficient to capture the correct physics. Even for $H = 40$, the velocity data is accurate whereas the shear stress data starts to become imprecise. Due to the similarity of the data for $H = 100$ and $H = 80$, we believe that a resolution $H = 100$ is sufficient. In the following, we will only report results extracted from the simulations with $H = 100$.

\paragraph{Index of distortion and flow relaxation}

In Fig.\ \ref{fig:indexdistortion}, the numerically obtained indexes of distortion for the velocity, $I_u$, and the shear stress, $I_\sigma$, are presented as function of the distance $d$ from the constriction. The corresponding ranges of decay, $\lambda_u$ and $\lambda_\sigma$, are shown in Tab.\ \ref{tab:rangeofdecay} and Fig.\ \ref{fig:ranges}. It is obvious that the slopes of $I_u(d)$ and $I_\sigma(d)$ can be excellently described by simple exponentials with decay lengths $\lambda_u$ and $\lambda_\sigma$, respectively. This justifies the approximations in Eqs.\ (\ref{eq:exponential_u}) and (\ref{eq:exponential_s}) and the introduction of the decay lengths $\lambda_u$ and $\lambda_\sigma$.

For small Reynolds numbers ($\text{Re} = 0.1$ and $1$), the curves of $I_u$ and $I_\sigma$ hardly depend on $\text{Re}$, cf.\ Figs.\ \ref{fig:Re0.1} and \ref{fig:Re1}. This is a first hint that $\text{Re} = 1$ still is a good approximation for Stokes flow. A significant change in the slopes is visible for larger Reynolds numbers ($\text{Re} > 1$) which can be seen from Figs.\ \ref{fig:Re10} through \ref{fig:Re100}. This is related to the influence of inertia.

There are only small differences between the inlet and outlet curves of $I_u$ and $I_\sigma$ for small Reynolds numbers, i.e., the flow fields are nearly symmetric with respect to the regions before and behind the constriction, cf.\ Figs.\ \ref{fig:Re0.1} and \ref{fig:Re1}. This is another hint for the validity of the Stokes limit at $\text{Re} \leq 1$. For larger $\text{Re}$, the indexes of distortion towards the outlet are always larger than those towards the inlet, indicating that the constriction mainly influences the flow behind itself, cf.\ Figs.\ \ref{fig:Re10} through \ref{fig:Re100}. Obviously, the symmetry is broken due to the presence of inertia. This can also be seen in Fig.\ \ref{fig:stresslines} where examples of the spatial shear stress evolution along the $x$-axis are shown. For $\text{Re} = 0.1$, the curves are symmetric with respect to the center of the constriction, but for $\text{Re} = 100$, the asymmetry is clearly visible.

Analyzing the data shown in Fig.\ \ref{fig:indexdistortion}, it is obvious that the decay characteristics of the distortion of the velocity and the shear stress are similar if not identical, i.e., $I_u(d) \approx I_\sigma(d)$ and $\lambda_u \approx \lambda_\sigma$ for a given Reynolds number. Since the shear stress is related to the spatial derivatives of the velocity, this observation indicates that there is only one characteristic decay length both for the velocity and the shear stress.

The increase of the outlet values of $\lambda_u$ and $\lambda_\sigma$ with $\text{Re}$ is shown in Tab.\ \ref{tab:rangeofdecay} and Fig.\ \ref{fig:ranges}. Qualitatively, the behavior of $\lambda(\text{Re})$ can be understood from Eq.\ (\ref{eq:developmentlength}), defining the development length $L_D$ of the velocity in a pipe as function of the Reynolds number. Although the definitions of $\lambda_u$ and $\lambda_\sigma$ on the one hand and $L_D$ on the other hand are not equivalent, both describe the same physics, namely the relaxation behavior of the fluid as a function of the Reynolds number. In Stokes flow, Eq.\ (\ref{eq:developmentlength}) yields a constant development length which is also the case in Fig.\ \ref{fig:ranges}. There is a transition region for Reynolds numbers in the interval $[10 - 100]$ after which $L_D(\text{Re})$ becomes linear in $\text{Re}$. This linear region of $\lambda_u$ and $\lambda_\sigma$, however, has not been probed in our simulations.

From Tab.\ \ref{tab:rangeofdecay} we find that the ranges of decay $\lambda_u$ and $\lambda_\sigma$ towards the inlet are always about $0.2 H$, regardless of the Reynolds number. The interpretation is that inertia affects only the fluid inside and behind the constriction. The fluid approaching the constriction from the inlet experiences the presence of the constriction only by momentum diffusion, and the Reynolds number does not play a significant role. This can also be seen by comparing the velocity and shear stress fields at different $\text{Re}$ upstream of the constriction shown in Fig.\ \ref{fig:streamlines_01_100}. For $\text{Re} = 0.1$ and $100$, the regions before the constriction look similar, but there are pronounced differences downstream. Applied to blood flow, this means that the constriction cannot cause clotting in the upstream region.

\paragraph{Peak values of velocity and shear stress}

The previous discussions clearly show that non-linear effects become important at large Reynolds numbers. In Figs.\ \ref{fig:stresslines} and \ref{fig:accumulation}, we present additional simulation data for the velocity and the shear stress to support those observations.

The spatial evolution of the shear stress along the $x$-axis is shown in Fig.\ \ref{fig:stresslines}. Here, $z = 0$ is fixed and $y / H = 0.005$ (close to the central axis at $y = 0$) in Fig.\ \ref{subfig:lines_1}, $y / H = 0.125$ (halfway between central axis and constricted walls) in Fig.\ \ref{subfig:lines_2}, and $y / H = 0.245$ (close to the constricted walls at $y / H = 0.25$) in Fig.\ \ref{subfig:lines_3}. Obviously, the shear stress distribution is symmetric with respect to $x = 0$ for $\text{Re} = 0.1$. At higher Reynolds numbers, $\sigma(x)$ is asymmetric.

Another important observation is that the peak value of the shear stress close to the wall increases disproportionally fast with $\text{Re}$, cf.\ Fig.\ \ref{subfig:lines_3}. While the fluid velocity is zero at the walls and maximum in the bulk region, the shear stress reaches its maximum close to or at the walls in a typical hydrodynamic flow situation. However, the lattice nature of the LBM causes inaccuracies in the computation of the shear stress close to inclined or curved obstacles. In other words: The numerical error of the shear stress close to the wall is increased. To diminish this problem, we have computed the average of the shear stress on an interval about the maximum of the curve $\sigma / \bar \sigma$, cf.\ Fig.\ \ref{subfig:lines_3},
\be
\label{eq:peak_stress}
\sigma_{\text{max}} = \f{1}{H/8} \int_{x_1}^{x_2} \text{d}x\, \sigma(x)
\ee
with $x_2 - x_1 = H / 8$ and $(x_1 + x_2) / 2 = -0.285 H$. The averaging process reduces possible lattice artifacts and is taken as a measure for the peak shear stress in the constriction. This procedure does not necessarily limit the significance of the quantity $\sigma_{\text{max}}$ since proteins and cells passing the constriction do not instantaneously react on the local shear stress. In fact, also the time of exposure plays a role (which is equivalent to a finite distance along the path due to the advection velocity). This is well-known in stress induced hemolysis \cite{beissinger1987low}. The results for the averaged peak shear stresses $\sigma_{\text{max}}$ as a function of $\text{Re}$ are presented in Fig.\ \ref{subfig:acc_s}. In Fig.\ \ref{subfig:acc_u}, the maximum velocity $u_{\text{max}}$ on the centerline ($y = z = 0$) is shown as function of the Reynolds number. It is normalized by the characteristic velocity $\bar u$ to enable comparability of the results for different Reynolds numbers.

In the Stokes limit, $u_{\text{max}} / \bar u$ and $\sigma_{\text{max}} / \bar \sigma$ do not depend on $\text{Re}$ since non-linear effects are absent. The location where the fluid reaches its peak velocity $u_{\text{max}}$ is either at $x = 0$ (smaller $\text{Re}$) or behind the middle, $x > 0$, of the constriction (larger $\text{Re}$), cf.\ Fig.\ \ref{subfig:comp_u}. At larger Reynolds numbers, $u_{\text{max}} / \bar u$ decreases. This can be understood qualitatively by comparing the time scales for diffusion and advection. On the one hand, at small Reynolds numbers, a distortion in the fluid mainly propagates by diffusion, and advection is negligible. On the other hand, advection is dominant at large Reynolds numbers. Since the constriction is a localized perturbation at the lateral walls, it takes some time until it can affect the fluid in the vicinity of the centerline. This time can be estimated by the diffusion time scale
\be
t_D = \f{H^2}{8 \nu}
\ee
where $\nu = \eta / \rho$ is the kinematic viscosity. In this time, however, the fluid has already propagated by a characteristic distance $L_{DP} = \bar u t_D$ where $L_{DP} / H \propto \text{Re}$. If $L_{DP}$ is large with respect to the length of the constriction, the fluid leaves the constriction and starts to relax again before the fluid near the central axis is fully aware of the perturbation caused by the constriction. Hence, the centerline velocity does not as strongly increase during the passage through the constriction as for smaller Reynolds numbers. This is also the reason for the increase of  $\sigma_{\text{max}} / \bar \sigma$ with $\text{Re}$ in Fig.\ \ref{subfig:acc_s}. The volume flux of the fluid through any cross-section perpendicular to the $x$-axis has to be constant. Thus, the average fluid velocity must become larger inside the constriction. When the velocity near the centerline is not proportionally increased (which is the case at large $\text{Re}$), the fluid near the walls has to be faster to compensate. This, on the other hand, leads to a disproportionate increase of the shear stress near the walls. In fact, for $\text{Re} = 100$,  $\sigma_{\text{max}} / \bar \sigma$ is about $53\%$ larger than in the viscous limit. This is a significant inertia effect which will be even more severe at $\text{Re} > 100$. The implication is that unfavorable blood vessel geometries in combination with large Reynolds numbers can lead to a significant non-linear build-up of shear stress causing further complications during stress-induced blood clotting.


\section{Conclusions}
\label{sec:conclusions}

Large shear stresses in blood flow can lead to a conformation change of the protein von Willebrand factor. This may trigger undesired blood clotting in arteries which can eventually lead to a coronary thrombosis. In order to estimate the impact of inertia on the shear stress in coronary arteries, we have employed the lattice Boltzmann method to simulate the flow in a constricted geometry with Reynolds numbers between $0.1$ and $100$. We assume the fluid to be Newtonian since the particulate nature of blood and its non-Newtonian properties are only significant in small blood vessels like venules and arterioles.

The major observation is that the peak value of the effective von Mises stress $\sigma_{\text{vM}}$ grows disproportionally fast with the Reynolds number in the inertial regime, $\text{Re} \geq 10$. At $\text{Re} = 100$, a common value of the Reynolds number in coronary arteries, the peak value of $\sigma_{\text{vM}}$ is more than $50\%$ larger than expected from assuming the validity of Stokes flow. This observation indicates that a combination of pathological blood vessel geometries and large Reynolds numbers may increase the risk of an heart attack. This is a pure hydrodynamic effect.

We further observe that the influence of the constriction is noticeable only inside and behind itself, i.e., upstream of the constriction, the flow field and the shear stress are not significantly influenced. The downstream distortion decays exponentially with the distance to the constriction, and its range grows linear with the Reynolds number for large $\text{Re}$. In particular, the inertial effects break the symmetry of the flow field upstream and downstream of the constriction.

With this article, we point out that pure hydrodynamic effects could be the reason for an increased tendency to blood clotting in pathologically altered blood vessel geometries in combination with large Reynolds numbers.


\begin{acknowledgements}
This project has been supported by the DFG grant VA205/5-1.
\end{acknowledgements}

\end{document}